# Ultra-fast re-structuring the electronic landscape of transparent dielectrics: new material states (Die-Met)

E.G. Gamaly and A.V. Rode

*Laser Physics Centre, Research School of Physics and Engineering, The Australian National University, Canberra ACT 0200, Australia*

e-mail: Eugene.gamaly@anu.edu.au; andrei.rode@anu.edu.au

**Abstract**

The swift excitation of the transparent dielectrics by intense short laser pulse produces ultra-fast re-structuring of the electronic landscape generating a wealth of material states continuously changing in space and in time in accord with the variations of the intensity. These unconventional transient material states combine simultaneously dielectric and metal properties (Die-Met). The laser excitation transforms a transparent dielectric into electrically inhomogeneous state early in the pulse time [1]. The permittivity of excited material varies in time and in space changing from positive to negative values that strongly affects the interaction process. The interplay between the transient permittivity gradient and polarisation of the incident laser becomes the major process of the new interaction mode. In a particular relation between the polarization and the permittivity gradient the incident field amplitude grows up while the wave is approaching to the surface where the real part of permittivity turns to zero. That results in the local increase in the absorbed energy density. The complex 3D structure of the permittivity makes a transparent part of excited dielectric (at $\varepsilon_0 > \varepsilon_{re} > 0$) optically active. The electromagnetic wave passing through such a medium gets a twisted trajectory and accrues the geometric phase [2]. The plane of polarisation rotation and phase depends on the 3D permittivity structure [3]. Measuring the polarisation and phase of the probe beam allows quantitatively identify this new transient state. We discuss the revelations of this effect in different experimental situations and possible applications.

## I. Introduction

The experimental and theoretical studies demonstrated that the swift excitation of the transparent dielectrics by short and intense laser pulses generates a wealth of continuously changing material states in space and in time in accord with the intensity of the beam.

The number of conduction electrons in the initially transparent dielectric grows up under the action of intense laser beam. The excited dielectric gradually transforms into a state where the forbidden band gap shrinks and the contribution of valence electrons into optical properties is lessening. At some absorbed fluence (and at corresponding number of conduction electrons) the contributions of conduction and valence electrons into polarization cancel each other and the real part of the permittivity turns to be zero. At that moment the running wave converts to the evanescent one. The unusual transient states of excited solid where dielectric and metallic properties are coexistent are generated lasting well after the pulse end. Note that the state when the real part of dielectric permittivity is zero is a special point for the Maxell equations.

The excitation produces ultra-fast re-structuring of the electronic landscape while the ions positions remain close to the initial states during several picoseconds needed for the energy transfer from the electrons to ions.

The transient states are differing by the band gap width (and allegedly with the different excited energy levels inside a forbidden gap), changed inter-atomic potential and optical properties. The spatial distribution of the transient dielectric permittivity becomes three-dimensional in accord with the spatial distribution of the laser field inside the interaction region. The real part of the excited permittivity passes through zero value becoming negative at some surface surrounding the plasma region in the centre of the beam. The presence of the zero-value surface significantly affects the laser-matter interaction process [4-6]. The importance and unusual properties of the state with zero permittivity (epsilon near zero, ENZ state) was recognized and studied in meta-materials in the microwave frequency domain [7]. Thus the laser-excited medium becomes electrically inhomogeneous offering another area for studies and applications. The pump-probe technique with time/space resolution makes in principle able to detect such states measuring the probe polarisation rotation along with the phase changes. May be in near future it will become possible to make some states metastable.

In what follows we discuss how states with mixture of dielectric and metallic properties are created including the zero-permittivity point (ZPP) state. Then we discuss some properties of these states and possible applications. The formation of the excited permittivity gradients including ZPP drastically changes the interaction mode. Finally we discuss possible experiments allowing to revealing some properties of these transient states and possible applications.

## II. Dielectric permittivity of the short laser-excited transparent dielectric

Transient dielectric permittivity of the laser-excited transparent dielectric can be presented in the approximate form similar to that of the Drude-like approach but with two major differences [1]. In the initially transparent dielectric the optical properties are dominated by the polarisation of the valence electrons while the conduction band is empty and absorption is close to zero. Ionisation by the multi-photon absorption process at the elevated intensity transfers electrons from the valence band to the conduction band. Ionisation gives rise to several effects. First, increasing number of the conductivity electrons starts contributing into the negative polarisation, while the contribution of the bonded electrons diminishes. Increasing the conduction electrons population results in absorption due to the inverse Bremsstrahlung process via the electrons collisions. Thus the collision rate grows up in in proportion to the conduction electrons number density. Below the approximation used in [1] is presented.

$$\varepsilon = \varepsilon_{re} + i\varepsilon_{im} \approx \varepsilon_0 - (\varepsilon_0 - 1)\frac{n_e}{n_a} - \frac{n_e}{n_{cr}(1 + v^2/\omega^2)} + i\frac{n_e}{n_{cr}(1 + v^2/\omega^2)}\frac{v}{\omega} \qquad (1)$$

Note that the above formula gives correct permittivity values in the limits of initial state and in plasma state. Collision rate by the kinetic definition reads:

$$v \approx n_e \cdot \sigma \cdot v_e$$

The rate depends on continuously growing the electron number density and electrons' velocity while the collision cross section might be approximately considered equal to atomic cross section, $\sigma \approx 10^{-15}$ cm$^2$, being practically constant during the ionization process. The collision rate grows up to the natural maximum, which is defined by the ratio of the electrons' velocity to the inter-atomic distance and comprises ~5x10$^{15}$ s$^{-1}$ [8]. Reasonable estimate for the energy of electrons transferred to the conduction band based on consideration that the electron energy is no less than that at the low band edge, $\varepsilon_{el} \approx \Delta_g$. Correspondingly $v_{el} = (2\Delta_g/m_e)^{1/2}$. However, one should take into account that with increasing the number of the conduction electrons the band gap shrinks. At the ionisation threshold the number density of the conduction electrons is around several percent. Thus, it is reasonable to suggest that the band gap decrease is small and the above estimate is appropriate. For sapphire one gets, $\Delta_g$ = 9.8 eV, $v_{el}$ = 2×10$^8$ cm/s. Then, $\nu \approx 2 \cdot 10^{-7} (cm^3/s) \cdot n_e$. The transient permittivity depends solely on the electron number density, which in turn depends on the local laser intensity. Therefore time/space map of the transient permittivity in laser-excited dielectric reproduces the time/space pattern of intensity inside the interaction region.

It is clear from (1) that there are two interaction regimes while the laser intensity increases during the pulse. First regime corresponds to the intensity range where the real part of the permittivity is positive. The electro-magnetic wave propagates through partly transparent excited crystal. In the second regime at the elevated intensity the real part of the permittivity turns to be negative. At the surface where $\varepsilon_{re}$ = 0 the running wave transforms to the evanescent wave absorbing in a medium converted to plasma. In what follows we discuss the transient states of matter generated in the first stage of the interaction.

### III. Transient material states generated at intensities when $\varepsilon_{re} \geq 0$

Ionization of a crystal starts with the threshold-free multi-photon process. This process firstly creates the excited states and afterwards electrons are transferred from intermediate state to the valence band [9]. Therefore at relatively low intensity the excited electron states are created. Continuously growing the conduction electrons number then affects the band gap structure. At some intensity the real part of the permittivity turns to zero when the positive polarisation of the valence electrons is cancelled by the negative contribution of the conduction electrons. Supposedly then the band gap collapses. The valence and conduction bands overlap after the first full ionization, i.e. conversion to plasma.

Allegedly the band gap between valence and conduction zone decreases when electrons from the valence zone are transferred into the conduction zone. Experimental results on the dependence of the band gap in silicon on the conduction electrons number density support the above assumption: band gap in silicon decreases almost two times (from 1.1 eV to 0.6 eV) while electron number density increases up to 10$^{18}$ cm$^{-3}$ due to heating. Of course, in the wide band gap dielectrics the electrons can be transferred to conduction zone only by ionization.

The decrease in the band gap should affect both multi-photon ionization rate and impact ionization as well. Keldysh accounted for the ionization from the excited states in the calculations of MPI ionization probability [9]. It is possible estimate the effect of the band gap decrease on impact ionization in classical approximation ($\Delta \gg \hbar\omega$).

Potential landscape is continuously changes in accord with time/space dependence of intensity in the interaction region. Growing number of the conductivity electrons gradually changes the inter-atomic potential. Indeed, the attractive part of the potential decreases due to transfer of valence electrons into the conduction band in the absence of electron-to-ion energy transfer. Therefore one cannot exclude the occurrence of intermediate potential minima corresponding to unknown metastable states.

Allegedly lifetime of the excited states is limited by the time period necessary for the energy transfer from electrons to ions. For laser-excited sapphire the energy transfer time is around 30-40 picoseconds.

Note that the transient states created by short laser pulses were observed in Gallium [10] Bismuth [11] and at the surfaces of atomic nanowires [12]. Transient real and imaginary parts of permittivity in short–laser-excited Bismuth measured by the pump/double-probe technique appeared to be different from those in the conventional solid and liquid. Lifetime of this transient Bismuth state amounted to four nanoseconds [11].

Summing up, the transient material states possess different permittivity, shrinking band gap and excited electronic levels within, modified inter-atomic potential. It is important to notice that the excited crystal remains transparent when $\varepsilon_0 > \varepsilon_{re} > 0$.

### IV. Reaching the zero permittivity state: Ionization threshold:

One can find with the help of (1) the number density of conduction electrons necessary to achieve the state where the real part of the permittivity is zero: $n_{\varepsilon_{re}=0} \approx \varepsilon_0 \cdot n_{cr}$. Critical density at 800 nm ($\omega = 2.356 \times 10^{15} s^{-1}$) equals to $n_{cr} = m_e c^2 \pi / e^2 \lambda^2 = 1.17 \cdot 10^{13} / \lambda^2 = 1.828 \times 10^{21}$ cm$^{-3}$. Thus for sapphire ($\varepsilon_0 = 3.06$) one gets, $n_{\varepsilon_{re}=0} = 5.598 \times 10^{21}$ cm$^{-3}$. Therefore in the state when real part of permittivity is zero in sapphire the number density of valence electrons constitutes 0.2% from those in the valence band. Electrons collide with neutrals, electrons and ions. The collision rate for the momentum exchange is around $\nu \approx 10^{15}$ s$^{-1}$. One immediately calculates the imaginary part of the permittivity, $\varepsilon_{im} = \dfrac{\varepsilon_0 \nu}{\omega(1+\nu^2/\omega^2)} = 1.1$; refractive index, $n = \kappa = (\varepsilon_{im}/2)^{1/2} = 0.74$; and absorption coefficient, $A = \dfrac{4\kappa}{(\kappa+1)^2 + \kappa^2} = 0.828$.

The laser flunce corresponding to the ionization threshold from condition that all absorbed energy density confined solely in the thermal energy of electrons transferred to the conduction band.

$$F_{\varepsilon_{re}=0}(t_{\varepsilon_{re}=0}) = \dfrac{3c \cdot n_{\varepsilon_{re}=0} \Delta_g}{4A(n_{\varepsilon_{re}=0}) \cdot \omega \cdot \kappa(n_{\varepsilon_{re}=0})} = 0.136 \text{ J/cm}^2$$

Thus hundred femtosecond long pulse (100 fs = $10^{-13}$ s) converts sapphire to strongly absorbing medium at intensity $1.36 \times 10^{12}$ W/cm$^2$. Running wave transforms to the evanescent wave. e-fold decrease length of the evanescent wave in sapphire at the ionization threshold: $l_{skin} = c/\kappa\omega = 165 nm$. Therefore the absorbed energy density equals to $8.24 \times 10^3$ J/cm$^3$. For comparison the enthalpy of formation for Aluminum oxide is 1,675.7 kJ/mol = $3.34 \times 10^5$ J/cm$^3$. The thermal energy at the melting point (2,345 K) equals to $1.16 \times 10^4$ J/cm$^3$. One may expect that the transient states might be reversible. The time dependence of the excited permittivity in intense 100 fs pulse is presented at Fig.1.

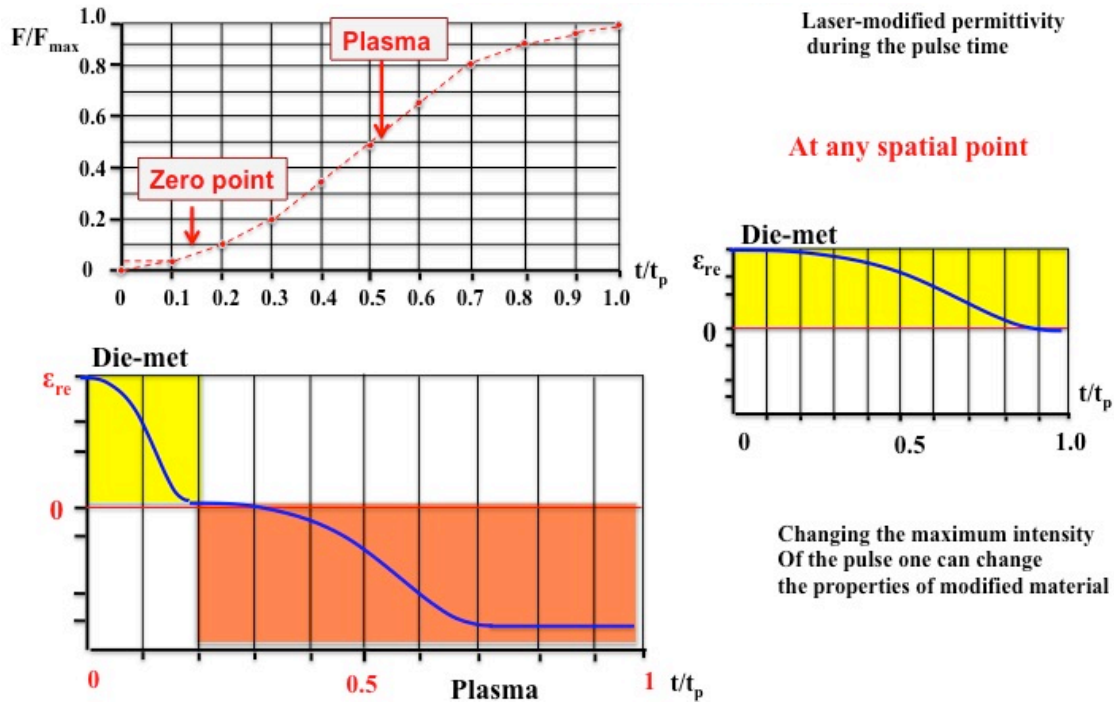

Fig.1 Time dependence of laser-modified permittivity and fluence for the Gauss pulse shape in time

**V. Spatial and temporal distributions of the transient states in laser-excited dielectric**

Let's consider the situation when the short laser pulse (the Gauss distribution in time and space) is normally incident on the transparent crystal surface. The maximum intensity at the center of the focal spot is equal to that necessary for achieving condition, $\varepsilon_{re} = 0$. Let us calculate the spatial distribution of the optical properties of excited crystal across the focal spot. In the table below we present optical properties as function of the electron number density. Afterwards with the help of the rate equation dependence of the electron number density on the intensity is established. Then for any spatial intensity distribution one can recover the spatial distribution of the excited optical properties in a crystal. At Fig.2 snapshot of the spatial distribution of excited permittivity is presented for the Gauss intensity distribution in space.

Table 1.

| intensity | low | | | ~ $10^{13}$ W/cm$^2$ |
|---|---|---|---|---|
| $\varepsilon_{re}$ | $\varepsilon_0$ | $\varepsilon_0 - 0.5$ | $\varepsilon_0 - 1$ | 0 |
| $n_e$ | negligible | $0.5 \times n_{cr}$ | $n_{cr}$ | $\varepsilon_0 \times n_{cr}$ |
| $\kappa$ | ~ 0 | 0.0108 | 0.047 | 0.74 |
| $l = c/\kappa\omega$ | transparent | 11.75 microns | 2.7 microns | 0.172 micron |

In the lower row is the length where the electric field e-fold decreases. One can see that at the critical density absorption (collision rate) is very low. It means that the dominant ionization process is MPI. Rate equation for the MPI with power dependence on intensity can be integrated analytically with the Gauss pulse time dependence (see appendix). The correspondence between intensity, electron number density, and optical properties, is then explicit.

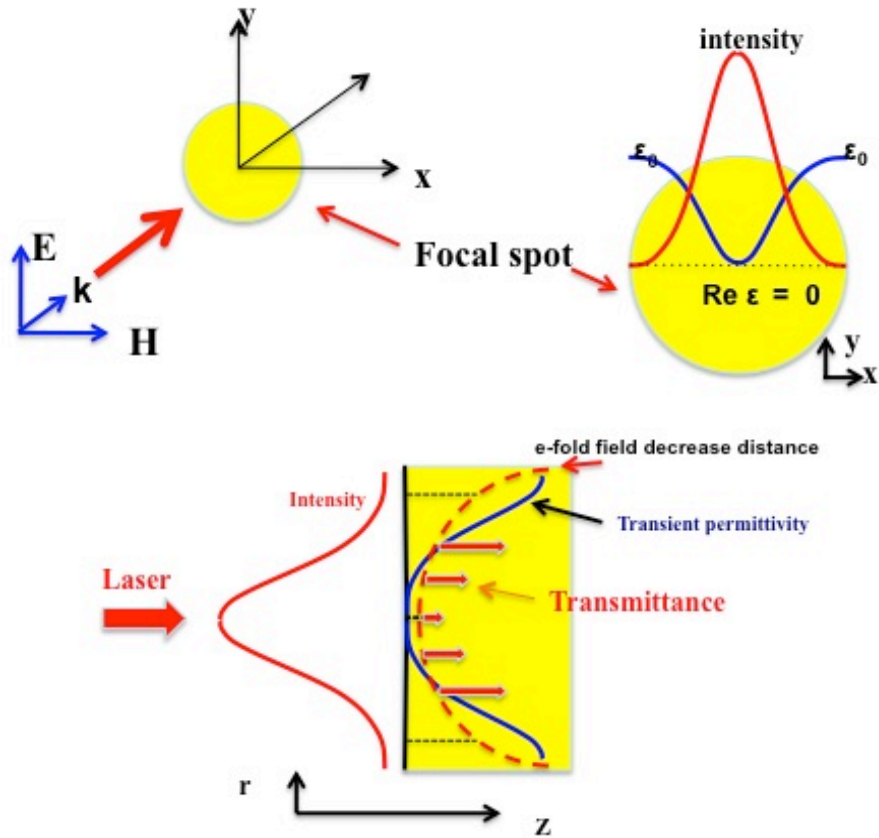

Fig.2. Snapshot of the spatial distribution of the optical properties in excited dielectric

Therefore, one can establish such a correspondence for a different spatial intensity distribution, for example for the Bessel beam focussing in a transparent dielectric (see Fig.3). In general the spatial distribution of the excited permittivity is three-dimensional and can be calculated following the above recipe. Therefore during the intense laser/ transparent dielectric interaction process laser-affected medium becomes electrically inhomogeneous: light at some moment begins interaction with a medium possessing the permittivity gradient where the real part of the permittivity changes from positive to negative values. It appears that the mode of interaction dramatically changes, which we consider in the following section.

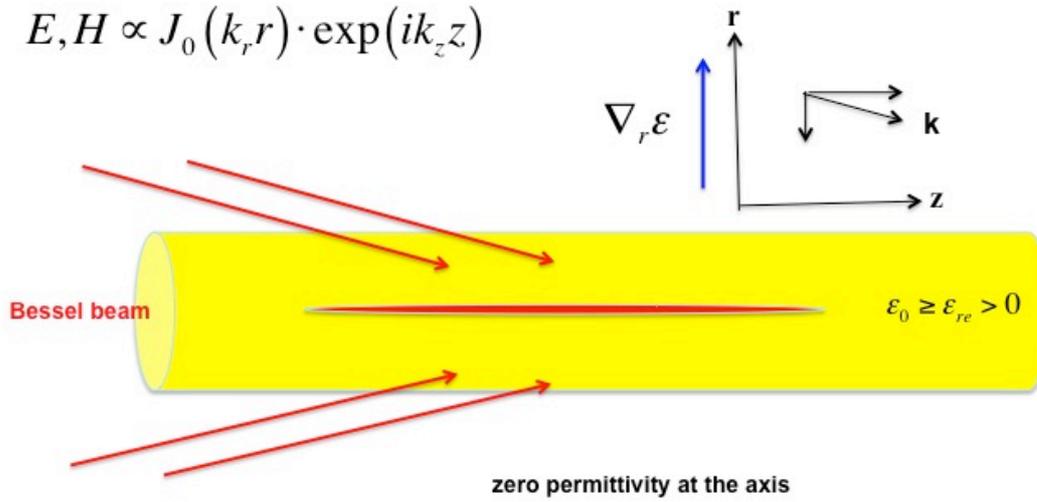

Fig.3. Scheme of the Bessel beam/transparent dielectric interaction

## VI. Light interaction with electrically inhomogeneous medium: permittivity gradient, polarization and zero permittivity point

It was found long ago that the electromagnetic wave interaction with the electrically inhomogeneous dielectric (the mass density remains homogeneous) has significant differences from the interaction with the isotropic medium. Let's consider the electromagnetic wave with the k-vector in x,z-plane incident on a layer in which the permittivity changes in z-direction [4]. There are two possible polarization directions: $H = H_y$; $E$ ($E_x$, $E_z$); and $E_y$; $H$ ($H_x$, $H_z$). The interaction does not depend on y-coordinate. K. Forsterling found [5] that wave with polarisation, $E$ ($E_x$, $E_z$), propagating through the inhomogeneous non-absorbing layer has sharp increase in the field amplitude (logarithmic singularity) near the point $\varepsilon = 0$. Later on Gildenburg [6] demonstrated that the infinitesimal absorption kills the singularity. However, the significant increase of the electric field occurs near the zero real part of the permittivity. The wave with polarization, $E$ ($E_x$, $E_z$), absorbing in the layer creates the energy flux in direction of the permittivity gradient inside the negative permittivity region. Energy dissipating in the unit volume is inverse proportional to the permittivity gradient.

The existence of singularity in the interaction of the light with inhomogeneous non-absorbing dielectric is inherent property of the Maxwell equations. It can be seen if the Maxwell equations are converted to the equation for the magnetic field [4]:

$$\Delta H - \frac{\varepsilon}{c^2}\frac{\partial^2 H}{\partial t^2} + \frac{1}{\varepsilon}\nabla\varepsilon \times (\nabla \times H) = 0$$

The last term in the above equation might be infinite at zero permittivity if the numerator in this term is non-zero. It is clear that the magnitude of this term depends on the relation between the permittivity gradient and filed polarisation as in the above considered case. The estimate for the field increase near the real part of the permittivity zero point might be obtained from the Maxwell equation. Indeed, from

$$\varepsilon \frac{\partial E}{\partial t} = \nabla \times H$$

one can get a reasonable estimate for the field amplitude increase in the vicinity of $\varepsilon_{re} = 0$, $E_{max}/E_0 = \left(\frac{\varepsilon_0}{\varepsilon_{im}}\right)$. Correspondingly intensity increases as $I_{max}/I_0 = \left(\frac{\varepsilon_0}{\varepsilon_{im}}\right)^2$. For sapphire the intensity increases almost 10 times in comparison with that in the incident wave (see Fig.4).

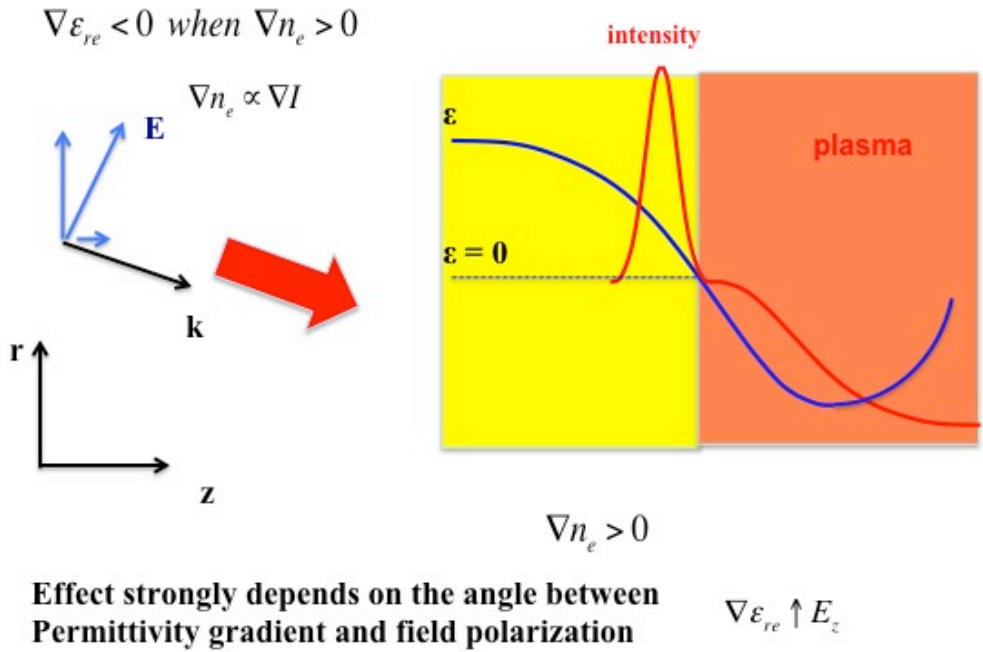

Fig.4 Electric field increase in the vicinity of the zero permittivity point

This effect might be especially significant for the Bessel beams interactions with transparent dielectrics due to big length of the focus and high intensity in the axial spike.

### VII. Observing the transient properties of the laser-excited dielectric by the probe beam

The spatial distribution of the real part of the transient permittivity across the focal spot in the excited dielectric changes from the positive values at the outer boundary of the beam to the negative values in the center of the focal spot for the Gaussian intensity distribution. In the circular ring surrounding the central plasma area real part of the permittivity changes from positive value of unexcited dielectric to zero. Zero value surface corresponds to the boundary between the excited, but still transparent, dielectric and that converted to plasma. Note that in this ring the band gap width decreases along with the decreasing permittivity in direction to the center of the beam.

It is reasonable suggesting that the excited states are created inside the initially forbidden gap during the band gap reduction.

Therefore interaction of the probe beam with this part of the excited dielectric may result in unusual phenomena such as a lasing if some excited state correspond population inversion for the probe beam photons or it might be a lasing without inversion.

The change in the real part of the refractive index during and after the excitation may be detected via the phase shift of the probe. An increase in the real part of the refractive index results in a positive phase shift whereas a decrease leads to a negative phase shift. The number of electrons in the conduction band is less than percent at the considered conditions. Therefore the number of empty energy levels largely exceeds the number of the occupied states. In these conditions the distribution function of electrons is close to the Maxwell-Boltzmann distribution [13].

The laser-excited dielectric is a medium possessing many properties allowing to be detected by the probe beam transmission through the laser-affected layer with time and space resolution. The value of the changed permittivity especially in the transparent wings of the focal spot might be measured by the interferometry of the transmitted beam and unperturbed beam. Another possibility relates to the measuring the transmitted beam plane of polarization rotation. Plane of polarization rotation in nonmagnetic medium is possible when:

1. Electric field has a component in the direction of the permittivity gradient (see Fig.5).

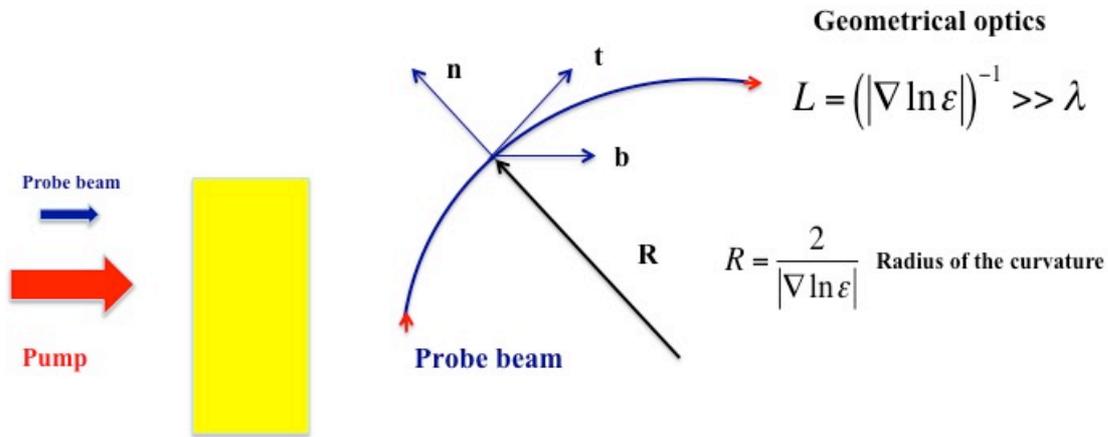

Fig. 5. Probe propagation through electrically inhomogeneous medium. Polarization rotates if the electric field has a component in the direction of the permittivity gradient.

In geometric optics approximation (using the Fermat's principle) the radius of the curvature of the light beam is inversely proportional to the permittivity gradient [3,4] (see Fig.5):

$$R = \frac{1}{\nabla \ln \varepsilon}$$

2. Medium is anisotropic then the light beam curve has torsion (see Fig.6). Let's denote arc length by $s$, unit tangent vector by $t$, principal normal vector by $n$, binormal vector by $b$. From the differential geometry follows: **b = t×n** . Radius of torsion is σ. Electric vector lies in normal (**b/n**) plane. During rotation electric vector makes an angle φ with normal. While moving along the beam the direction of the electric vector (polarization) rotates in accord to equation [2,4]:

$$\frac{d\varphi}{ds} = \frac{1}{\sigma}$$

Thus, measuring the angle of the polarization rotation for the beam passing through the layer of thickness L allows to define the radius of torsion. Differential geometry relations allow connecting the radius of torsion to the radius of curvature of the probe beam trajectory inside an excited layer and therefore obtain the average permittivity gradient value (see Fig.6).

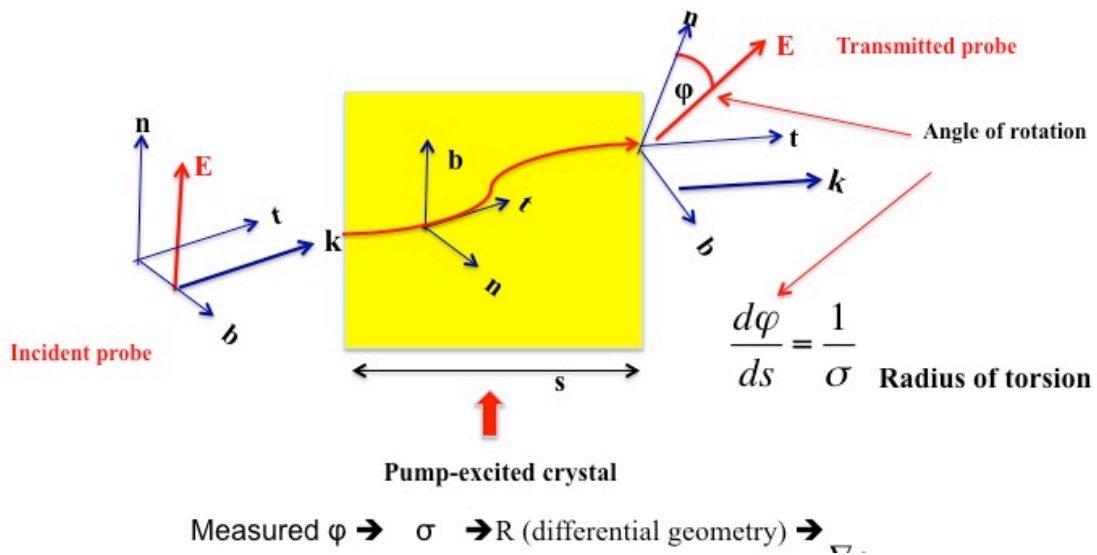

Fig.6. Plane of polarization rotation in an anisotropic medium (geometrical phase). The probe propagates along the twisted (non-planar) trajectory. The electric vector rotates in *b,n* plane

Using the probe beam with the photon energy less than the band gap (300-400 nm, energy 3-5 eV) allows in principle to catch up with some excited levels inside a band gap or with the absorbing edge of the shrinking band gap.

## VII. Conclusions, future directions

The interaction of short intense laser beam with transparent dielectric produces continuously changing electronic landscape in space and in time while the ions positions remain close to unperturbed state until the energy transfer from excited electrons to ions.

The transient sates created possess with decreased in comparison to the initial state permittivity, shrinking band gap with allegedly excited levels, decreased binding energy and shifted minima in the inter-atomic potential. The range of these parameters can be controlled by variation of laser intensity in space and time.

At sufficiently elevated intensity of the pump laser the real part of the permittivity changes from positive to negative values, which signifies transformation to plasma state. Interaction of the incident beam with the surface of zero permittivity in electrically inhomogeneous medium results in the sharp field increase near the zero point when the field polarization is directed along the permittivity gradient. It seems to be most important in the interaction of the Bessel beam due to the elongated focal area. The laser excitation at intensity below ionization threshold ($\varepsilon_{re} \geq 0$) converts dielectric into electrically anisotropic medium. Measuring reflection, transmission, plane of polarization rotation, phase shift of the probe beam propagating through laser excited layer might reveal many properties of these transient states possessing simultaneously with metal and dielectric properties.

At this stage of studies it is difficult to predict what might be the possible applications of these unusual states. Ultra-fast sensors or detectors might be a straightforward application. If it would be possible to stabilize some transient state intermediate between initial and Zero Permittivity Point by multiple pulse action then novel metastable material phase might be uncovered.

The future directions of these studies heavily depend on the results of pump/probe experiments, which may reveal the properties of these states.

## Acknowledgements

We acknowledge funding from the Australian Government through the Australian Research Council's Discovery Projects funding scheme (DP170100131).